# Sorting Methods and Adaptive Thresholding for Histogram Based Reversible Data Hiding

**Ammar Mohammadi [1] and Mansor Nakhkash [1]***

[1] Department of Electrical Engineering, Yazd University, Yazd, Iran
*Corresponding author e-mail: nakhkash@yazd.ac.ir

**Abstract:** This paper presents a histogram based reversible data hiding (RDH) scheme, which divides image pixels into different cell frequency bands to sort them for data embedding. Data hiding is more efficient in lower cell frequency bands because it provides prediction that is more accurate. Using pixel existence probability for some pixels of ultra-low cell frequency band, another sorting is performed. Employing these two novel sorting methods in combination with the hiding intensity analysis that determines optimum prediction error, we improve the quality of the marked image especially for low embedding capacities. In effect, comparing to existent RDH algorithms, the hiding capacity is increased for a specific level of the distortion for the marked image. Experimental results confirm that the proposed algorithm outperforms state of the art ones.

## 1 Introduction

Data hiding is to embed data in a cover medium such as audio, multimedia, image, and textual files. Embedding data in image has many applications, such as copyright protection, authentication and cover communication. The original cover image in many applications of data hiding is not required to be restored. However, there exist applications in medical image sharing, multimedia archive management and so on, which need the restoration of the cover without any distortion. Reversible data hiding (RDH) is an approach that addresses the methods for complete restoration of the original cover after the extraction of embedded data.

There are various RDH methods [1]; but the more developed methods can be classified into two main categories: the methods based on 1- DE (difference expansion) [2] and 2- histogram modification [3]. In addition to this grouping, the two main DE and histogram based techniques can be combined with prediction-error expansion [4-11], lossless compression [12-17], code division multiplexing [18] and prediction-error modification [19-26].

The difference expansion (DE) was proposed by Tian [2] in 2003. In DE, the image is divided into pairs of pixels, in which the data bits are embedded. This method embeds one bit into a pixel pair and thus a hiding rate up to 0.5 bit per pixel can be realized. Tian exploits a location map to record the selected expandable locations, where the difference of pixel pairs is small enough to make less distortion. Location maps, usually, are enormous in size occupying a notable portion of the payload. Kamstra et al. [27] outperform the method of [2] by sorting pixel pairs according to the local variance. The basic idea of sorting is used by several schemes proposed later. In 2007, Thodi and Rodrguez [4] exploited histogram shifting technique and prediction-error (PE) expansion to improve DE scheme. Applying histogram shifting technique, they eliminate the need for location map and employ information of neighboring pixels to obtain an efficient PE for expanding. Reference [4] improves the scheme in [27] especially for higher embedding capacities (EC)s. In 2009, Sachnev et al. [5] employed rhombus predictor to embed data into an image. Their scheme divides a cover image into two sets denoted as cross and dot. Half of the data bits are embedded in cross (dot) set and the other half is embedded in dot (cross) set. Their method is better than the one in [27] because they use sorting in terms of local variance with histogram shifting and cross-dot embedding mechanism. In addition, the use of rhombus predictor makes sharper error histogram that is an improvement to the scheme of [4].

Modifying pairs of peak and zero points in the image histogram, Ni et al. [3] present an innovative RDH scheme. The pixels between peak and zero points of histogram are modified to make space for embedding data. Lee et al. [19] exploit histogram modification of difference image to embed data. There is a large probability that neighboring pixels in an image have similar pixel values. That means the histogram of differences between two neighboring pixels is sharper than the histogram of the original image. Thus, it provides higher peak point resulting in higher capacity to embed data. It is better to name the scheme in [19] as PE modification, which indicates three operations: predict a pixel, calculate PE and modify PE to embed data bits. In 2007, Fallahpour and Sedaaghi [28] improved the scheme in [3] by executing it on series of blocks. Almost, all schemes, proposed later on the basis of the histogram, use the idea of PE modification. Yang and Tsai [20], later on, introduced a histogram based RDH scheme, in which they used rhombus predictor [5] to embed data. It employs PE modification to embed data bits. Li et al. [21] introduce a RDH scheme based on two-dimensional difference histogram modification and difference-pair-mapping. They use gradient-adjusted-prediction (GAP) predictor to make difference-histogram. The employment of two-dimensional difference-histogram results in more embedding capacity that reduces the number of shifted pixels and distortion. Li et al. [22], also, propose multiple histogram modification as a new embedding mechanism. They use cross-dot embedding mechanism and rhombus predictor, which are introduced in [5]. Instead of the local variance in [5] and [27], complexity measurement is used in their method to construct PE histogram. Their algorithm has great performance and outperforms the methods in [5] and [21]. Ma and Shi [18] propose a RDH algorithm based on code division multiplexing. They employ Walsh Hadamard matrix to generate orthogonal spreading sequences, in which one can embed overlapped data bits without interfering. They, also, employ cross-dot embedding mechanism and rhombus predictor. Their scheme can be considered as a kind of histogram modification that has great performance; but not at low EC. In 2017, Wang et al. [23] introduced a histogram-based method and an organized framework to design multiple RDH scheme with nearly optimal rate and distortion performance. Their scheme is independent of cover images and uses genetic algorithm to realize optimization. In 2019, Xiao et al. [24] introduced a content dependent pairwise embedding scheme. In order to embed data, they modify two-dimensional histogram of PEs that is divided into specific regions. The selection of the expansion bins in these regions is formulated as an optimal path searching problem to adaptively modify PEs.

This paper provides three proposals to improve the efficiency of a RDH method. This improvement can be viewed as either to decrease the distortion for a specific number of embedding bits or to increase the embedding capacity for a specific degree of distortion. The first one is a sorting method based on dividing image pixels into cell frequency bands from the lowest to highest one. In this way, the pixels are ranged from smooth to rough. The second proposal is another sorting and concerned with dividing the cell frequency bands into some sub-bands by the use of existence probability of some pixels in the lowest band. Employing the two sorting algorithms, one starts embedding from the smoothest pixels towards the roughest and, therefore, the low distortion can be achieved. Hiding intensity analysis is our third proposal to adaptively determine optimum thresholds for a rhombus predictor RDH method in [20]. The implementation of our proposals and comparing the results with those of the existent RDH algorithms indicate the improvement in PSNR for a fix number of embedding bits. In another view, the embedding capacity is increased considerably for a specific PSNR.

## 1.1 Related Work

One of the best predictors in RDH schemes is rhombus predictor [5]. It is introduced in [20] as chessboard (CB) predictor (Fig. 1) as well. In this predictor, black pixels are predicted using white neighbors and vice versa. Let $h_{i,j}$ denotes the intensity of $i,j$th pixel in a cover image H. Four pixels in the set $\{h_{i-1,j}, h_{i+1,j}, h_{i,j-1}, h_{i,j+1}\}$ that we call a cell, are used to hide one data bit in $h_{i,j}$. To realize hiding in [20], $h_{i,j}$ can be predicted by its four neighbors as

$$h^p_{i,j} = \lfloor \frac{h_{i,j-1} + h_{i+1,j} + h_{i,j+1} + h_{i-1,j}}{4} \rfloor \qquad (1)$$

where $\lfloor . \rfloor$ is the round function. Using $h^p_{i,j}$ and $h_{i,j}$, the prediction-error $e_{i,j}$ is computed by

$$e_{i,j} = h_{i,j} - h^p_{i,j} \qquad (2)$$

for each intensity.

The bit $b$ that can be 0 or 1, changes the error according to two thresholds $SV_p \geq 0$ and $SV_n < 0$ in histogram of the PE. Fig. 2 shows a sample histogram of PE that we employ to describe how to insert data bits. The serviceable values (SVs) are those, in which the data bits are embedded. Shifting the intensity value of some pixels in two directions, the unserviceable values (USVs) are formed to make space and perform hiding. The right shift and left shift involve the increase and decrease of pixel intensities that are at the right side of a positive threshold $SV_p$ and at the left side of a negative threshold $SV_n$ respectively. The change of error is accomplished as [20]

$$e'_{i,j} = \begin{cases} e_{i,j} + 1 & \text{if } e_{i,j} > SV_p \\ e_{i,j} - 1 & \text{if } e_{i,j} < SV_n \\ e_{i,j} + \text{sign}(e_{i,j})\, b & \text{if } e_{i,j} = SV_p \text{ or } e_{i,j} = SV_n \\ e_{i,j} & \text{else} \end{cases} \quad (3)$$

where sign indicates the signum function. Embedding the data, the intensity of $i,j$th pixel $(m_{i,j})$ in the marked image M is given by

$$m_{i,j} = h^p_{i,j} + e'_{i,j} \quad (4)$$

After transmitting the marked image to a destination, data extraction and cover image reconstruction are carried out as follows. The recipient computes $e'_{i,j}$ by

$$e'_{i,j} = m_{i,j} - h^p_{i,j} \quad (5)$$

and then retrieves $e_{i,j}$ and extracts bit $b$ using (6).

$$e_{i,j} = \begin{cases} e'_{i,j} - 1 & \text{if } e'_{i,j} > SV_p \\ e'_{i,j} + 1 & \text{if } e'_{i,j} < SV_n \\ e'_{i,j}, b = 0 & \text{if } e'_{i,j} = SV_p \text{ or } e'_{i,j} = SV_n \\ e'_{i,j} - 1, b = 1 & \text{if } e'_{i,j} = SV_p + 1 \\ e'_{i,j} + 1, b = 1 & \text{if } e'_{i,j} = SV_n - 1 \\ e'_{i,j} & \text{else} \end{cases} \quad (6)$$

Finally, the cover image is reconstructed as

$$h_{i,j} = h^p_{i,j} + e_{i,j} \quad (7)$$

It should be noted that $h^p_{i,j}$ can be computed at destination from $\{m_{i-1,j}, m_{i+1,j}, m_{i,j-1}, m_{i,j+1}\}$ using (1), in which $h$ is replaced by $m$. The CB predictor in [20] can be started from the prediction for either black or white pixels. Starting from the cells that contain four black pixels for the prediction, data bits are embedded according to Eq. (1), (2) and (3) in white pixels. After embedding a number of data bits in the white pixels, embedding continues in black ones by the prediction that is calculated from the cells with four neighboring white pixels, and this process can be iterated to embed all data bits.

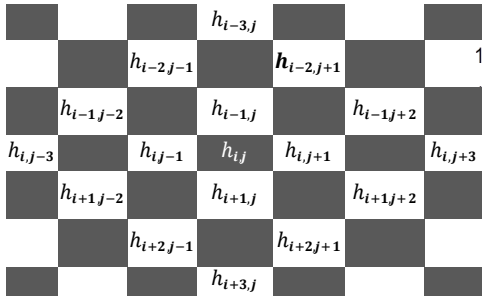

Fig. 1 A part of an image, whose pixels are divided into white and black sections.

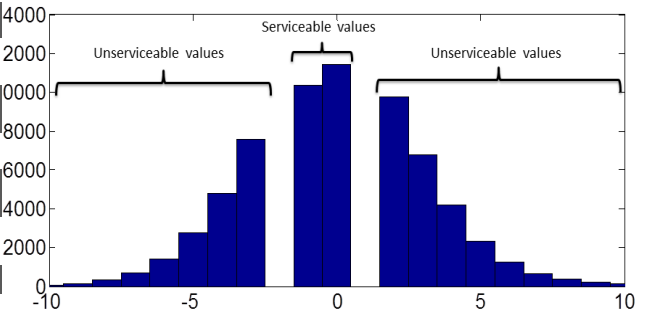

Fig. 2 Serviceable and unserviceable values in the histogram of PEs.

## 2 Proposed Scheme (Individual Descriptions)

Our proposal comprises two new sorting methods and an algorithm to determine adaptively the thresholds $(SV_p, SV_n)$ in the embedding method described in Section 1.1. They are introduced and explained as follows.

### 2.1 The Sorting Method Using Cell Frequency Spectrum

On the basis of local complexity, reference [7] suggests to divide image pixels into smooth and rough groups and the bits are hided in smooth and rough regions in an adaptive manner. Also in [22], some thresholds are exploited to distinguish smooth image regions from rough ones. This paper introduces the concept of cell frequency bands to distinguish image regions. The cell frequency bands are ranged from ultra-low frequency (ULCF) to ultra-high frequency (UHCF) bands. The rationale behind this concept is that the ULCF band defines the smoothest region, where the sharpest PE histogram is obtained for pixel intensities. On the contrary, the roughest region is specified by UHCF band. Thus, the embedding procedure is started from the pixels of ULCF band to achieve the least distortion for the original image. In the following, the separation of the frequency bands are explained.

As mentioned in Section 1.1, a cell comprises four pixels that are neighbor to a target pixel. Data bits are embedded in target pixels and neighboring ones remain intact. We determine cell smoothness or cell roughness using local difference (LD) of the neighboring pixels, and then classify cells to different cell frequency bands. For the target pixel with intensity $h_{i,j}$ in Fig. 1, neighboring white pixels $\mathbf{H}_{center} = \{h_{i-1,j}, h_{i+1,j}, h_{i,j-1}, h_{i,j+1}\}$ are sorted from small to large intensity and the sorted set is denoted as $\{S1, S2, S3, S4\}$. In order to classify a cell, a local difference $LD_{center}$ is computed according to

$$LD_{center} = (|S4 - S3| + |S4 - S2| + |S4 - S1| + |S3 - S2| + |S3 - S1| + |S2 - S1|)/6 \qquad (8)$$

It can be understood when $LD_{center}$ is low, the intensities of four neighboring pixels are close together and hence, the cell specifies a smooth region. Because $S4 > S3 > S2 > S1$, (8) can be simplified to

$$LD_{center} = (S4 - S1)/2 + (S3 - S2)/6 \qquad (9)$$

Equation (9) shows the allocation of more weight to the difference of side pixels (*S4-S1*) than middle ones (*S3-S2*). Neighboring side cells $(\mathbf{H}_{left}, \mathbf{H}_{right}, \mathbf{H}_{low}, \mathbf{H}_{up})$ in addition to $\mathbf{H}_{center}$ are, also, exploited to better determine the smooth or the rough region of an image. The side cells include three pixels instead of four pixels, i.e. $\mathbf{H}_{right} = \{h_{i+1,j+2}, h_{i,j+1}, h_{i-1,j+2}\}$, $\mathbf{H}_{left} = \{h_{i,j-1}, h_{i+1,j-2}, h_{i-1,j-2}\}$, $\mathbf{H}_{up} = \{h_{i-2,j-1}, h_{i-1,j}, h_{i-2,j+1}\}$ and $\mathbf{H}_{low} = \{h_{i+2,j-1}, h_{i+1,j}, h_{i+2,j+1}\}$ (Fig. 1). If the three pixels in the side cells are sorted from small to large intensities and their rearrangement is called $\{I1, I2, I3\}$, we can compute $LD$ for the side cells as

$$LD_{side} = (|I3 - I2| + |I3 - I1| + |I2 - I1|)/3 = 2(I3 - I1)/3 \qquad (10)$$

**TABLE 1** Cell frequency spectrum

| Thresholds | Cell Band |
|---|---|
| $f_c \leq T_{ulcf}$ | ULCF |
| $T_{ulcf} < f_c \leq T_{vlcf}$ | VLCF |
| $T_{vlcf} < f_c \leq T_{lcf}$ | LCF |
| $T_{lcf} < f_c \leq T_{mcf}$ | MCF |
| $T_{mcf} < f_c \leq T_{hcf}$ | HCF |
| $T_{hcf} < f_c \leq T_{vhcf}$ | VHCF |
| $T_{vhcf} < f_c$ | UHCF |

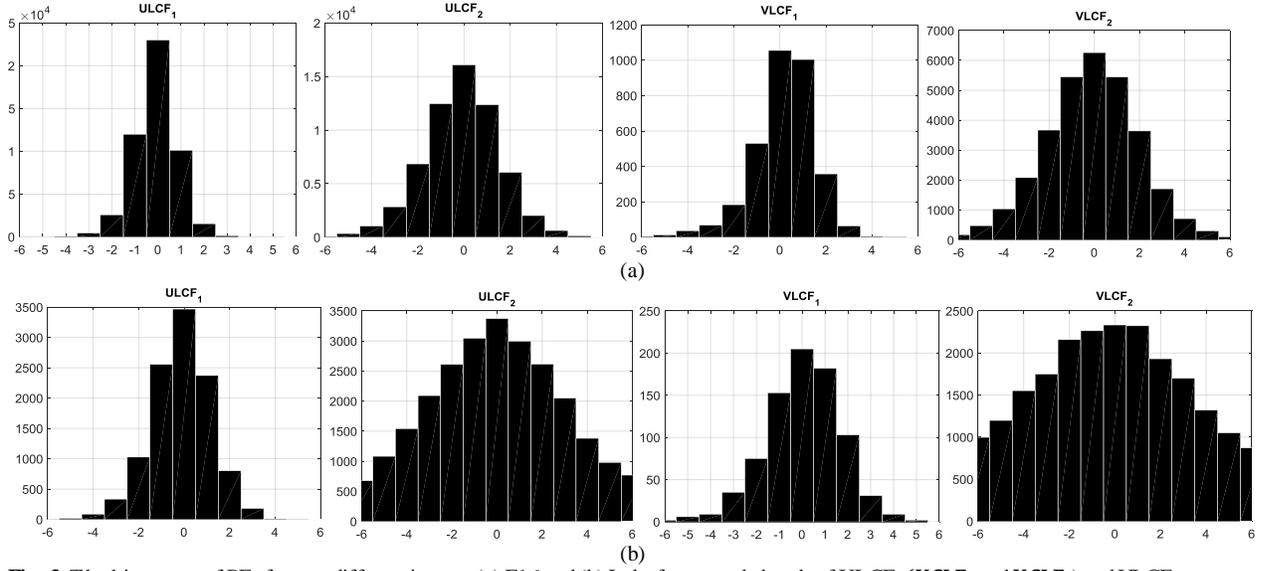

**Fig. 3** The histogram of PEs for two different images (a) F16 and (b) Lake for two sub-bands of ULCF (**UCLF**$_1$ and **UCLF**$_2$) and VLCF (**VCLF**$_1$ and **VCLF**$_2$) bands respectively.

The quantity $LD$ for all four side cells are calculated using (10) and are denoted as $LD_{right}, LD_{left}, LD_{up}$ and $LD_{low}$. The cell frequency classification is accomplished according to the mean value of the local differences.

$$f_c = \overline{LD} = (LD_{center} + LD_{right} + LD_{left} + LD_{up} + LD_{low})/5 \tag{11}$$

In Table 1, with regard to $f_c$ and thresholds $T_{ulcf}, T_{vlcf}, T_{lcf}, T_{mcf}, T_{hcf}$ and $T_{vhcf}$, the cells are classified to ultra-low cell frequency (ULCF), very low cell frequency (VLCF), low cell frequency (LCF), medium cell frequency (MCF), high cell frequency (HCF), very high cell frequency (VHCF) and ultra-high cell frequency (UHCF). The bands with lower cell frequency are more desirable and make less distortion for data embedding. Therefore, the data is embedded in a cover image according to the sorting of Table 1.

## 2.2 The Sorting Method Based on Pixel Existence Probability (SPEP)

The ULCF band, for its sharper histogram of PE, is the best band for data embedding. This subsection describes another sorting, in which the ULCF band is divided into some sub-bands. This sorting is based on the existence probability of pixels in a part of ULCF with lower LD that we call it L_ULCF. In order to describe the procedure for sub-banding ULCF, VLCF and etc. bands, let $\mathbf{O}_\hbar$ indicates the set of pixels for the original image with intensity $\hbar$ and $\mathbf{U}_\hbar$ shows the set of pixels with the same intensity $\hbar$ that belongs to L_ULCF. In addition, let $\mathcal{N}_\hbar$ and $\mathcal{M}_\hbar$ indicate the number of pixels in $\mathbf{O}_\hbar$ and $\mathbf{U}_\hbar$ respectively. Because we have $\mathbf{U}_\hbar \subset \mathbf{O}_\hbar$, if a pixel $x_\hbar$ is selected randomly from $\mathbf{O}_\hbar$, the probability that this pixel being in $\mathbf{U}_\hbar$ is obtained from (12).

$$f_\hbar = P(x_\hbar \in \mathbf{U}_\hbar | x_\hbar \in \mathbf{O}_\hbar) = \frac{\mathcal{M}_\hbar}{\mathcal{N}_\hbar} \tag{12}$$

$f_\hbar$ is called the existence probability of the pixel $x_\hbar$ in L_ULCF and it is calculated by (12) for $\hbar = 0$ to 255, i.e. $\mathbf{F} = \{f_0, \dots, f_\hbar, \dots f_{255}\}$ is the set of the existence probabilities for all intensities.

A smooth area owns two specifications: 1- pixel intensities in this area have low variation 2- There are a considerable number of pixels in a smooth area. The L_UCLF embodies pixels with the sharpest histogram of PE in the ULCF band and hence, it satisfies the first specification. In order to specify an area with considerable number of pixels in the L_UCLF band (the second specification), we obtain the intensity $\hbar_0 = \arg\max_{0 \leq \hbar \leq 255}(f_\hbar)$ that has the most relative papulation in the L_UCLF. Defining $U_{\hbar_0}$ as a set of pixels with intensity $\hbar_0$ in the UCLF, set $U_{\hbar_0}$ is more probable to be the smoothest areas and the data embedment should be started from the pixels in $U_{\hbar_0}$. Reasoning in the same way, the data embedment should be continued with pixels having the intensity $\hbar_1$, $\hbar_1 = \arg\max\limits_{\substack{0\leq \hbar \leq 255 \\ \hbar \neq \hbar_0}}(f_\hbar)$ and so forth.

To implement the SPEP procedure in ULCF band, we arrange the members of set **F** from large to small, and denote it as $\mathbf{F}_S = \{f_{\hbar_0}, \ldots, f_{\hbar_m}, \ldots, f_{\hbar_{255}}\}$. The class $\mathbf{U}_S = \{\mathbf{U}_{\hbar_0}, \mathbf{U}_{\hbar_1}, \ldots, \mathbf{U}_{\hbar_m}, \ldots, \mathbf{U}_{\hbar_{255}}\}$ expresses the pixel sets in ULCF band corresponding to $\mathbf{F}_S$. Set $\mathbf{U}_{\hbar_0}$ includes the smoothest areas and $\mathbf{U}_{\hbar_{255}}$ involves the roughest ones. In order to reduce the computational complexity of the implementation, we can divide $\mathbf{U}_S$ into some subclasses and perform data hiding for each subclass. Thus, we specify $L$-1 thresholds $f_{\hbar_{t1}}, f_{\hbar_{t2}}, \ldots, f_{\hbar_{t(L-1)}}$ and divide $\mathbf{U}_S$ to $L$ subclasses as $\mathbf{U}_S = \{\mathbf{ULCF}_1, \mathbf{ULCF}_2, \ldots, \mathbf{ULCF}_L\}$. The subclasses are $\mathbf{ULCF}_1 = \{\mathbf{U}_{\hbar_0}, \mathbf{U}_{\hbar_1}, \ldots, \mathbf{U}_{\hbar_{t1}}\}$, $\mathbf{ULCF}_2 = \{\mathbf{U}_{\hbar_{t1+1}}, \mathbf{U}_{\hbar_{t1+2}}, \ldots, \mathbf{U}_{\hbar_{t2}}\}$, $\ldots$, $\mathbf{ULCF}_L = \{\mathbf{U}_{\hbar_{t(L-1)+1}}, \mathbf{U}_{\hbar_{t(L-1)+2}}, \ldots, \mathbf{U}_{\hbar_{255}}\}$. Starting data hiding from $\mathbf{ULCF}_1$ to $\mathbf{ULCF}_L$, we can reduce the computational complexity of the implementation with slight loss of quality. The classification can be viewed as the split of ULCF band into $L$ sub-bands.

Although we can divide all bands, e.g. LCF, MCF and etc., to different sub-bands, it is more efficient to apply the division just to ULCF and VLCF bands. That is because there is a trade-off between computational complexity and marked image quality. It should be, also, noted that the set $\mathbf{F}_S$ is only obtained from L_ULCF once and it is used to sub-band all ULCF, VLCF and etc. bands. That means if pixels in ULCF band with intensity $\hbar_0$ define a smooth area, pixels in VLCF with the same intensity $\hbar_0$ specify a smooth area as well.

We illustrate the division of UCLF and VLCF to sub-bands by an example. Fig. 3 shows the PE histograms of F16 and Lake images for two different sub-bands in ULCF and VLCF bands, respectively, ($\mathbf{ULCF}_1$ and $\mathbf{ULCF}_2$) and ($\mathbf{VLCF}_1$ and $\mathbf{VLCF}_2$). In this example, 3000 pixels of ULCF band, which have lower LD, are chosen as L_ULCF to bring out $\mathbf{F}_S$. The ULCF and VLCF bands are divided into two sub-bands, which are classified according to $f_{\hbar_{t1}} = 0.01$ and $f_{\hbar_{t1}} = 0.05$, respectively, for F16 and Lake. In F16 image, the pixels $x_h$ with the intensities in interval $210 < \hbar < 225$ have $f_\hbar \geq 0.01$ and are classified as $\mathbf{ULCF}_1$ and $\mathbf{VLCF}_1$ sub-bands. We subsequently choose other pixels in these bands to construct $\mathbf{ULCF}_2$ and $\mathbf{VLCF}_2$ (Fig. 3a). For Lake image, the threshold is set to $f_{\hbar_{t1}} = 0.05$ that specifies interval $220 < \hbar < 231$ for $\mathbf{ULCF}_1$ and $\mathbf{VLCF}_1$ (Fig. 3b).

As it is demonstrated for both F16 and Lake images, the PEs belong to $\mathbf{ULCF}_1$ sub-band have the sharpest histogram, which means these pixels are smoother and more predictable than $\mathbf{ULCF}_2$ ones. Also, $\mathbf{VLCF}_1$ provides sharper histogram than $\mathbf{VLCF}_2$ for PEs. It is concluded that $\mathbf{ULCF}_1$ and $\mathbf{VLCF}_1$ are more suitable to embed data bits than $\mathbf{ULCF}_2$ and $\mathbf{VLCF}_2$ respectively. In practice, instead of pixel own intensities, (1) is employed to obtain the predicted intensity of pixels in the implementation of SPEP method.

### 2.3 Hiding Intensity Analysis

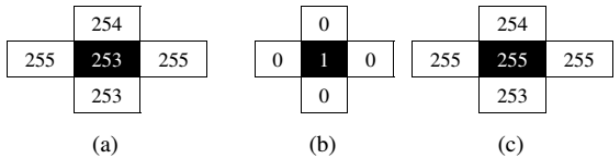

**Fig. 4** Classification of three black pixels to green pixel ($C_g$) ((a) and (b)) and yellow one ($C_y$) (c).

**Algorithm 2: Classifying pixels to $C_y$ or $C_g$**

if $255 - h_{i,j}^p < \tau_2$ and $h_{i,j} \geq h_{i,j}^p$ then
    $h_{i,j}$ is a $C_y$
else if $h_{i,j}^p < \tau_1$ and $h_{i,j} < h_{i,j}^p$ then
    $h_{i,j}$ is a $C_y$
else
    $h_{i,j}$ is a $C_g$
end if

**Algorithm 1: Adaptive determination of $SV = (SV_p, SV_n)$ according to HIA**

Do
if $N_b > N_{sv}(k_{poh})$ then
    $SV(k_{poh})$ is exploited
    $N_b = N_b - N_{sv}(k_{poh})$
    Use the next sub-band or band for the rest of bits
else
    for $k = 1:K$
        if $N_b/N_{USV} \leq HI_{sorted}(k)$ then
            $SV(k)$ is exploited
            $N_b = 0$
            **Break** For Loop
        end if
    end for
end if
while $N_b > 0$

To improve the efficiency in data embedding, we introduce hiding intensity analysis (HIA) that determines adaptively two thresholds $SV_p \geq 0$ and $SV_n < 0$ in CB predictor method [20]. As explained in Section 1.1, the serviceable and unserviceable values are to be specified according to two thresholds $SV_p$ and $SV_n$. The thresholds can be $(SV_p, SV_n) = (0, -1)$ or $(1, -2)$ or etc. The number of serviceable values ($N_{sv}$) and unserviceable ones ($N_{usv}$) depend on the pair $(SV_p, SV_n)$. $N_{usv}$ is the number of pixels distorted to make space for embedding bits. Therefore, if $N_{usv}$ is kept minimum for a specified number of embedding bits ($N_b$), the distortion for the marked image will be less. In this regard, we define a hiding intensity ($HI$) criteria that is a function of $(SV_p, SV_n)$ and given by

$$HI(SV_p, SV_p) = N_{sv}/N_{usv} \tag{13}$$

The $HI$ criteria shows how many bits can be embedded for each unserviceable pixels. Table 2a lists the aforementioned parameters for a sample image for different $(SV_p, SV_n)$.

With regard to the $HI$ criteria, the embedding should be started with the pair $(SV_p, SV_n)$, for which $HI(SV_p, SV_p)$ is maximum. However, if the number of hiding bits $N_b$ is more than the number of serviceable values $N_{sv}$ with maximum $HI$, it is required to reiterate embedding for the next frequency bands. Under this condition, the distortion to the marked image could be less if the embedding is started with a pair $(SV_p, SV_n)$, for which $N_{sv} \geq N_b$. In case the condition $N_{sv} \geq N_b$ cannot be satisfied for any $(SV_p, SV_n)$, the embedding should be started with the pair $(SV_p, SV_n)$, for which $N_{sv} = N_{sv}^{Max}$. It is noted that $N_{sv}^{Max}$ is the maximum number of serviceable values obtained for a specific $(SV_p, SV_n)$, e.g. $N_{sv}^{Max} = 21795$ for $(SV_p = 0, SV_n = -1)$ in Table 2a.

As a result, we introduce another criteria called power of hiding ($PoH$) and defined as

$$PoH = N_{sv} \times HI = N_{sv}^2/N_{USV} \tag{14}$$

The criteria $PoH$ is a compromise to emphasize both $N_{sv}$ and $HI$, and its maximum is used to develop our procedure for adaptive determination of $SV = (SV_p, SV_n)$. In order to describe the procedure, let us sort the rows of Table 2a according to $HI$ values from high to low and put the results in Table 2b. Also, let us obtain the row number corresponding to the maximum $PoH$ as

$$k_{poh} = \arg \max_{1 \leq k \leq K} PoH(k) \tag{15}$$

where the capital $K$ indicates the total number of rows. For an example, $k_{poh} = 5$ and $\max PoH = PoH(5) = 15371$ in Table 2b. With regard to the above explanation, the procedure to determine adaptively $SV = (SV_p, SV_n)$ for embedding bits is given by Algorithm 1.

## 2.4 Overflow and Underflow

It is possible for a marked image, whose some pixels get overflow or underflow in grayscale values. That means the grayscale values of some pixels in the marked image may exceed the upper bound (255 for an eight-bit grayscale image) or lessen the lower bound ("0") respectively. In [6], Luo et al. propose a useful method to overcome overflow or underflow (OU). In [6], after a level of data hiding, if there are 256 and -1, they will be modified to 255 and 0. In order to discriminate the modified 255 and 0 values from the original ones, they assign 1 to all modified pixels and 0 to the original ones. These 0 and 1, as overhead information, are compressed and inserted in the marked image. Using the overhead information, modified pixels will be recovered at the receiver.

**TABLE 2** Hiding intensity analysis for a sample image

| (a) | | | | | | (b) | | | | |
|---|---|---|---|---|---|---|---|---|---|---|
| Number | SV | $N_{SV}$ | $N_{USV}$ | HI | PoH | Number | SV | $N_{SV}$ | $N_{USV}$ | $HI_{sorted}$ | PoH |
| 1 | {-1,0} | 21795 | 43819 | 0.49 | 15371 | 1 | {-5,4} | 3734 | 4214 | 0.89 | 3515 |
| 2 | {-2,1} | 17384 | 26435 | 0.66 | 14097 | 2 | {-4,3} | 6963 | 7948 | 0.88 | 6517 |
| 3 | {-3,2} | 11524 | 14911 | 0.77 | 10131 | 3 | {-3,2} | 11524 | 14911 | 0.77 | 10131 |
| 4 | {-4,3} | 6963 | 7948 | 0.88 | 6517 | 4 | {-2,1} | 17384 | 26435 | 0.66 | 14097 |
| 5 | {-5,4} | 3734 | 4214 | 0.89 | 3515 | 5 | {-1,0} | 21795 | 43819 | 0.49 | 15371 |

We propose a novel algorithm to decrease or eliminate the need for overhead information in [6]. First, let's define three pixel groups, green ($C_g$), yellow ($C_y$) and red ($C_r$) pixels. $C_g$ is a pixel with no or very low probability of OU. $C_y$ is susceptible to OU, and $C_r$ is a pixel with OU, whose value is -1 or 256. Embedding is done on green pixels ($C_g s$), and $C_y s$ are ignored in the process of embedding because they pose more possibility to generate $C_r s$. Nevertheless, the existence of $C_r s$ must be controlled after data embedding and dealt with them by applying the method in [6].

A pixel is classified to $C_y$ or $C_g$ using Algorithm 2, where $\tau_1$ and $\tau_2$ are whole number that are employed to classify some pixels near lower bound or upper bound to $C_y$. Also, in this algorithm, $h_{i,j}^P$ is the prediction amount of $h_{i,j}$. It is calculated using (1). The greater $\tau_1$ and $\tau_2$ is selected, the more pixels are classified as $C_y$ and it leads to less probability for generating OU (i.e. generating $C_r$ pixels).

With an example (Fig. 4), we will explain Algorithm 2. Let's choose $\tau_1 = 2$ and $\tau_2 = 2$. Regarding Algorithm 2 and Fig. 4a, we have $h_{i,j}^P = 254$, $h_{i,j} = 253$ and $255 - h_{i,j}^P < \tau_2$ but $h_{i,j} \geq h_{i,j}^P$ is not satisfied; therefore, it will be classified as $C_g$. In this case, because $h_{i,j} < h_{i,j}^P$, $h_{i,j}$ either remains intact or decreases after embedding data (using (3) and (4)) and no overflow occurs. There is a similar argument for Fig. 4b, where $h_{i,j}^P = 0$, $h_{i,j} = 1$. For this cell, we have $h_{i,j}^P < \tau_1$ but the condition $h_{i,j} < h_{i,j}^P$ is not satisfied; hence, $h_{i,j}$ either remains intact or increases and no underflow occurs. The black pixel in Fig. 4b belongs to $C_g$ as well.

In Fig. 4c, considering $h_{i,j} > h_{i,j}^P$ ($h_{i,j}^P = 254$, $h_{i,j} = 255$), there is a possibility that $m_{i,j}$ overflows with respect to (3) and (4). Therefore, this pixel is classified as $C_y$ and will be ignored in the process of embedding.

It is possible to prevent an OU in $C_g s$ completely if proper values for $\tau_1$ and $\tau_2$ are selected. However, one should control whether any OU occurs after embedding data and deal with it using the method in [6].

## 3 Overall Proposed RDH Scheme

In this section, the whole proposed RDH scheme comprising the methods of data embedding, data extracting and retrieving the cover image are described. To embed data, the total image pixels are divided into two black and white pixel groups. Also, all bits of the data are divided into two equal parts of the black and the white ones to be, respectively, embedded in the black and white pixels. Data embedment can be begun with white or black ones. If the white bits are embedded in the white pixels at first, the black ones remain unchanged. In the next step, the black bits are embedded in the black pixels and the white ones remain intact. We refer to the whole embedment in white and black pixels as one embedding level.

After the end of hiding process, overhead information is inserted in some determined locations of the image, e.g. in the first or last row (or column), where has been remained intact during data embedding. The overhead information includes an indication whether the embedment begins with the white or black pixels, the total number of data bits, $SV$ values, the number of data hiding steps, and the OU information,

We illustrate the procedure of embedding data and lossless recovering of the original image after extracting data bits, respectively, in Fig. 5a and Fig. 5b. More details of the proposed scheme for embedding (Fig. 5a) are described

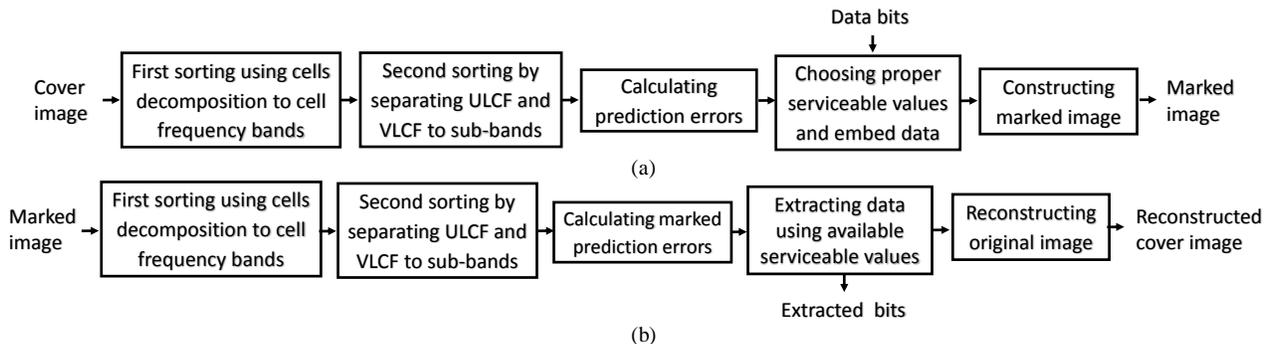

**Fig. 5** Sketch of the proposed RDH scheme for one level of (a) embedding and (b) extracting data and recovering the original image.

as follows :

1) First sorting: calculate $f_c$ from (11) for each cell. Cells are classified according to $f_c$ (Table 1) from ULCF to UHCF bands.
2) Second sorting: divide ULCF and VLCF bands into several sub-bands with respect to Section 2.2. Embedding is begun with **ULCF$_1$**.
3) Calculating PEs: Using (1), the values $h_{i,j}^p$ are obtained. PEs ($e_{i,j}$) in the selected frequency band or sub-band are calculated for all cells using (2).
4) Data embedding: according to Section 2.3, *HI* and *PoH* are calculated for different *SV* values; the best *SV* ($SV_n, SV_p$) will be picked up with respect to Algorithm 1 and the values $e'_{i,j}$ are calculated from (3). In this equation, we apply *SV* just for $C_g$s to control overhead information for OU (Section 2.4). Multi-level embedding: If all black bits are embedded, we start hiding white bits in white pixels from step 1. Otherwise, we continue embedding black bits in black pixels of the higher cell frequency bands. Also, after one embedding level, it will be continued for more levels to embed all data. After any embedding level, we must check for $C_r$. If there is $C_r$, we can use method that is presented in [6] to manage it.
5) Constructing marked image: the marked image is created using (4). After data embedding process, the overhead information is embedded in the marked image.

At the receiver, initially, the overhead information is extracted. Based on the number of data bits, *SV* values, hiding steps, starting from the black or white pixels and the OU information, the data extraction and original image retrieving are begun. Considering the hiding process has been started from black cells, extracting data bits and retrieving the cover image must be started from white ones. This process is explained as follows.

1) First sorting: classify the cells from ULCF to UHCF bands using neighboring pixels and the calculated $f_c$.
2) Second sorting: divide ULCF and VLCF bands into several sub-bands. Sort cells based on $f_c$.
3) Calculating marked PEs: $h_{i,j}^p$ is computed at destination from $\{m_{i-1,j}, m_{i+1,j}, m_{i,j-1}, m_{i,j+1}\}$ using (1), in which *h* is replaced with *m*. Then using (5), $e'_{i,j}$ is calculated.
4) Data extraction: using (6), $e_{i,j}$ is recovered for $C_g$s and data bits are extracted. It should be noted that ($SV_n, SV_p$) has been sent to the recipient as overhead information.
5) Recovering original image: pixel intensities for coordinate (*i*,*j*) of the original image is retrieved for the specified cell frequency band, using (7). It continues for complete white pixels.
Multi-level extracting and recovering: if all white data bits are extracted, we start to extract black bits from Step 1. Otherwise, we will keep on extracting the rest of white bits from the higher cell frequency bands or sub-bands. If the extraction reaches the UHCF band, we would go to the next level of extraction beginning with ULCF sub-bands in order to extract more bits and restore the original image.

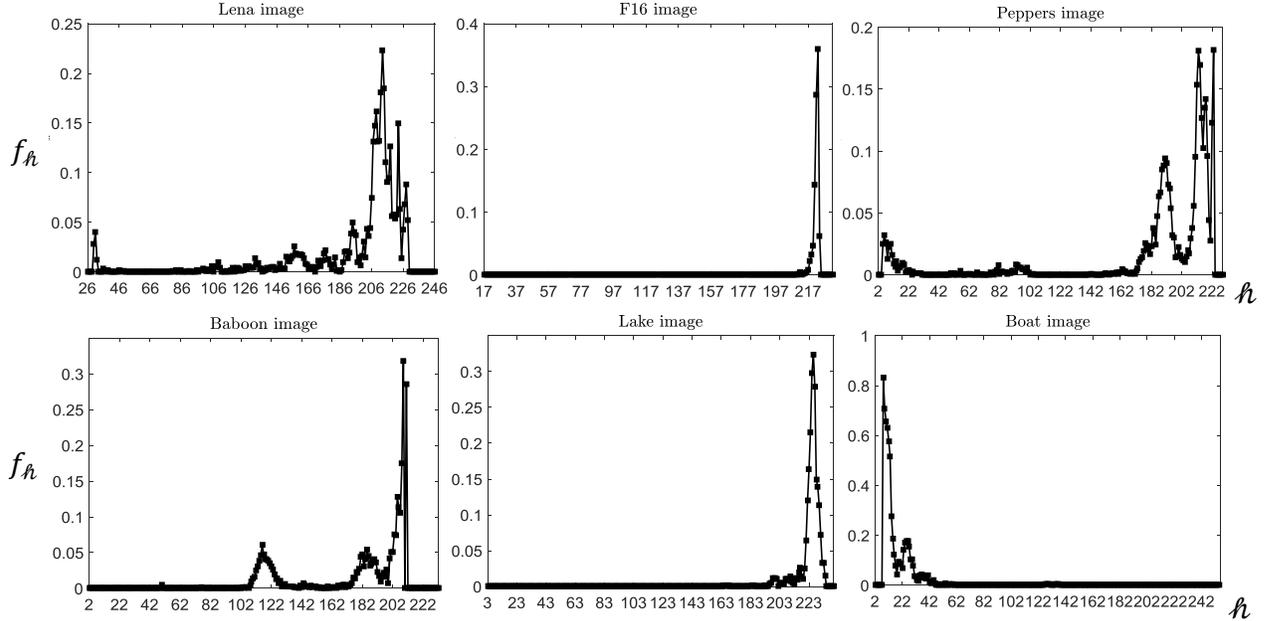

**Fig. 6.** Calculated $f_\hbar$ for intensity $\hbar$ for the test images.

## 4 Experimental Result

We have conducted several experiments to certify the performance of the proposed method in hiding capacity and image quality of the marked image. Six gray scale images, Lena, F16, Peppers, Baboon, Lake, and Boat all $512 \times 512$ in size are used as test images. We embed pseudo-random binary data in the test images. PSNR is employed as a measure for the marked image quality in these experiments. According to thresholds $T_{ulcf} = 3.3$, $T_{vlcf} = 4.5$, $T_{lcf} = 6$, $T_{mcf} = 9$, $T_{hcf} = 13$, and $T_{vhcf} = 18$, cells are divided into ULCF, VLCF, LCF, MCF, HCF, VHCF and UHCF bands (Table 1). In our experiments, the UHCF band that involves roughest cells is ignored. As stated before, if hiding is initially executed with white pixels, black ones as neighboring pixels will be remained intact. Hence, the next step of embedding is begun with black ones, while all the thresholds are increased by a small bias around 0.3 to update cell frequency bands. This bias accounts for the reduction of image smoothness after one embedding step.

In the Lena and F16 images, because of having no pixel intensity near the upper or lower bound, the OU can, just, occur for high EC. Lake image is subject to underflow. However, liable pixels to underflow are at the borders of the Lake image, e.g. pixels in row 512, that are not used to embed data bits in our implementation. Hence, there is no need to apply Algorithm 2 for these three images. In Peppers and Baboon the underflow is more possible due to the presence of pixel intensities near lower bound. We set $\tau_1 = 13$ and $\tau_1 = 37$ in Algorithm 2 for Peppers and Baboon, respectively, to avoid underflow for ECs up to $10^5$ bits. Moreover, for both ones $\tau_2$ is zero in Algorithm2. Therefore, Algorithm2 ignores 2470 and 959 pixels of Peppers and Baboon images respectively, i.e. they are considered as $C_y$.

There exist intensities near both lower and upper bounds in Boat image and thus, its pixels are susceptible to both underflow and overflow. These pixels are all in UHCF band that ignores in data hiding process and no need to apply

**TABLE 3** PSNR results of 6 test images for two $f_c$s and different ECs.

| | EC | 0.5 | 1 | 2 | 5 | 10 | 20 | 40 | 70 | 100 | $\times 10^3$ bits |
|---|---|---|---|---|---|---|---|---|---|---|---|
| **Lena** | $f_c = \overline{LD}$ | 74.51 | 71.04 | 67.43 | 62.78 | 59.27 | 55.81 | 52.23 | 48.59 | 45.63 | dB |
| | $f_c = LD_{center}$ | 72.50 | 69.50 | 66.42 | 62 | 58.63 | 55.34 | 52.03 | 48.41 | 45.53 | dB |
| **F16** | $f_c = \overline{LD}$ | 77.04 | 73.94 | 70.84 | 66.45 | 63.11 | 59.62 | 55.84 | 52.43 | 49.44 | dB |
| | $f_c = LD_{center}$ | 75 | 72.72 | 69.53 | 65.20 | 62.36 | 58.88 | 55.31 | 52.27 | 49.25 | dB |
| **Peppers** | $f_c = \overline{LD}$ | 69.98 | 66.61 | 63.45 | 59.18 | 55.91 | 52.55 | 48.66 | 43.75 | 40.58 | dB |
| | $f_c = LD_{center}$ | 69.13 | 66.12 | 63 | 58.72 | 55.41 | 52.31 | 48.54 | 43.64 | 40.46 | dB |
| **Baboon** | $f_c = \overline{LD}$ | 69.43 | 66.27 | 62.93 | 58.47 | 54.93 | 50.18 | 44.33 | 39.01 | 30.13 | dB |
| | $f_c = LD_{center}$ | 68.52 | 65.24 | 62.13 | 57.89 | 54.55 | 49.89 | 44.06 | 38.81 | 30.65 | dB |
| **Lake** | $f_c = \overline{LD}$ | 74.17 | 71 | 67.67 | 62.47 | 58.10 | 53.23 | 48.47 | 43.02 | 39.44 | dB |
| | $f_c = LD_{center}$ | 72.47 | 69.51 | 66.02 | 61.19 | 57.35 | 53.09 | 48.16 | 42.92 | 39.35 | dB |
| **Boat** | $f_c = \overline{LD}$ | 73.02 | 69.66 | 66.16 | 61.26 | 57.09 | 52.88 | 48.82 | 43.79 | 40.50 | dB |
| | $f_c = LD_{center}$ | 70.96 | 67.70 | 64.52 | 60.15 | 56.58 | 52.88 | 48.58 | 43.69 | 40.47 | dB |

**TABLE 4** The effect of HIA and SPEP tools to improve PSNR of 6 test images.

| | EC | 0.5 | 1 | 2 | 5 | 10 | 20 | 40 | 70 | 100 | $\times 10^3$ bits |
|---|---|---|---|---|---|---|---|---|---|---|---|
| **Lena** | HIA | 74.66 | 71.42 | 68.43 | 64.10 | 60.60 | 56.94 | 52.97 | 48.88 | 45.73 | dB |
| | SPEP | 74.65 | 71.35 | 68.14 | 63.86 | 59.63 | 55.82 | 52.24 | 48.62 | 45.63 | dB |
| | SPEP & HIA | 76.42 | 73.05 | 69.56 | 64.47 | 61 | 56.93 | 53.09 | 48.92 | 45.74 | dB |
| **F16** | HIA | 76.35 | 73.16 | 70.05 | 66.33 | 62.70 | 59.96 | 55.84 | 52.67 | 49.91 | dB |
| | SPEP | 77.62 | 73.95 | 70.85 | 66.63 | 63.32 | 59.94 | 55.85 | 52.43 | 49.45 | dB |
| | SPEP & HIA | 77.62 | 74.65 | 71.64 | 67.54 | 64.21 | 59.96 | 56.53 | 52.97 | 50.03 | dB |
| **Peppers** | HIA | 73.14 | 70.17 | 66.83 | 61.71 | 58.04 | 54.28 | 49.65 | 44.57 | 41.14 | dB |
| | SPEP | 70.26 | 66.87 | 63.57 | 59.30 | 55.99 | 52.54 | 48.69 | 43.79 | 40.59 | dB |
| | SPEP & HIA | 74.69 | 70.80 | 66.87 | 62.03 | 58.15 | 54.35 | 49.66 | 44.61 | 41.18 | dB |
| **Baboon** | HIA | 71.59 | 68 | 64.56 | 59.85 | 56.10 | 50.95 | 44.59 | 39.29 | 35 | dB |
| | SPEP | 69.46 | 66.34 | 62.94 | 58.49 | 54.94 | 50.18 | 44.34 | 39.17 | 30.14 | dB |
| | SPEP & HIA | 71.42 | 68.06 | 64.85 | 60.05 | 56.06 | 50.95 | 44.61 | 39.32 | 35 | dB |
| **Lake** | HIA | 72.65 | 69.35 | 65.77 | 62.22 | 58.10 | 53.80 | 48.93 | 43.45 | 39.80 | dB |
| | SPEP | 74.52 | 71.43 | 68.32 | 63.90 | 58.19 | 53.22 | 48.51 | 43.02 | 39.44 | dB |
| | SPEP & HIA | 76.72 | 73.54 | 69.83 | 63.86 | 58.82 | 53.64 | 49.1 | 43.45 | 39.80 | dB |
| **Boat** | HIA | 73.18 | 69.73 | 66.83 | 61.59 | 57.74 | 53.78 | 49.31 | 44.29 | 40.83 | dB |
| | SPEP | 73.13 | 70.01 | 66.69 | 61.53 | 57.24 | 52.88 | 48.88 | 43.79 | 40.51 | dB |
| | SPEP & HIA | 75.67 | 71.90 | 68.01 | 62.37 | 58.36 | 53.87 | 49.31 | 44.32 | 40.88 | dB |

Algorithm 2 to Boat image.

In the experiments, the ULCF and VLCF bands are divided just into two sub-bands. Fig. 6 shows probability $f_\hbar$ for the test images. We use 3000 pixels of ULCF band with lower LD as L_ULCF. According to this figure and choosing $f_{\hbar_{t1}} = 0.04, 0.01, 0.03, 0.05, 0.05$ and $0.02$ respectively for Lena, F16, Peppers, Baboon, Lake and Boat, we divide ULCF and VLCF bands into two sub-bands. With respect to these values for $f_{\hbar_{t1}}$, the pixels with intensities in ranges $202 < \hbar < 229, 210 < \hbar < 225, 183 < \hbar < 223, 196 < \hbar < 212, 220 < \hbar < 231$, and $7 < \hbar < 39$ are chosen in ULCF or VLCF as $\mathbf{ULCF}_1$ or $\mathbf{VLCF}_1$ sub-bands, respectively, for 6 aforementioned images. Other pixels in ULCF or VLCF can, therefore, be selected, respectively, as $\mathbf{ULCF}_2$ or $\mathbf{VLCF}_2$.

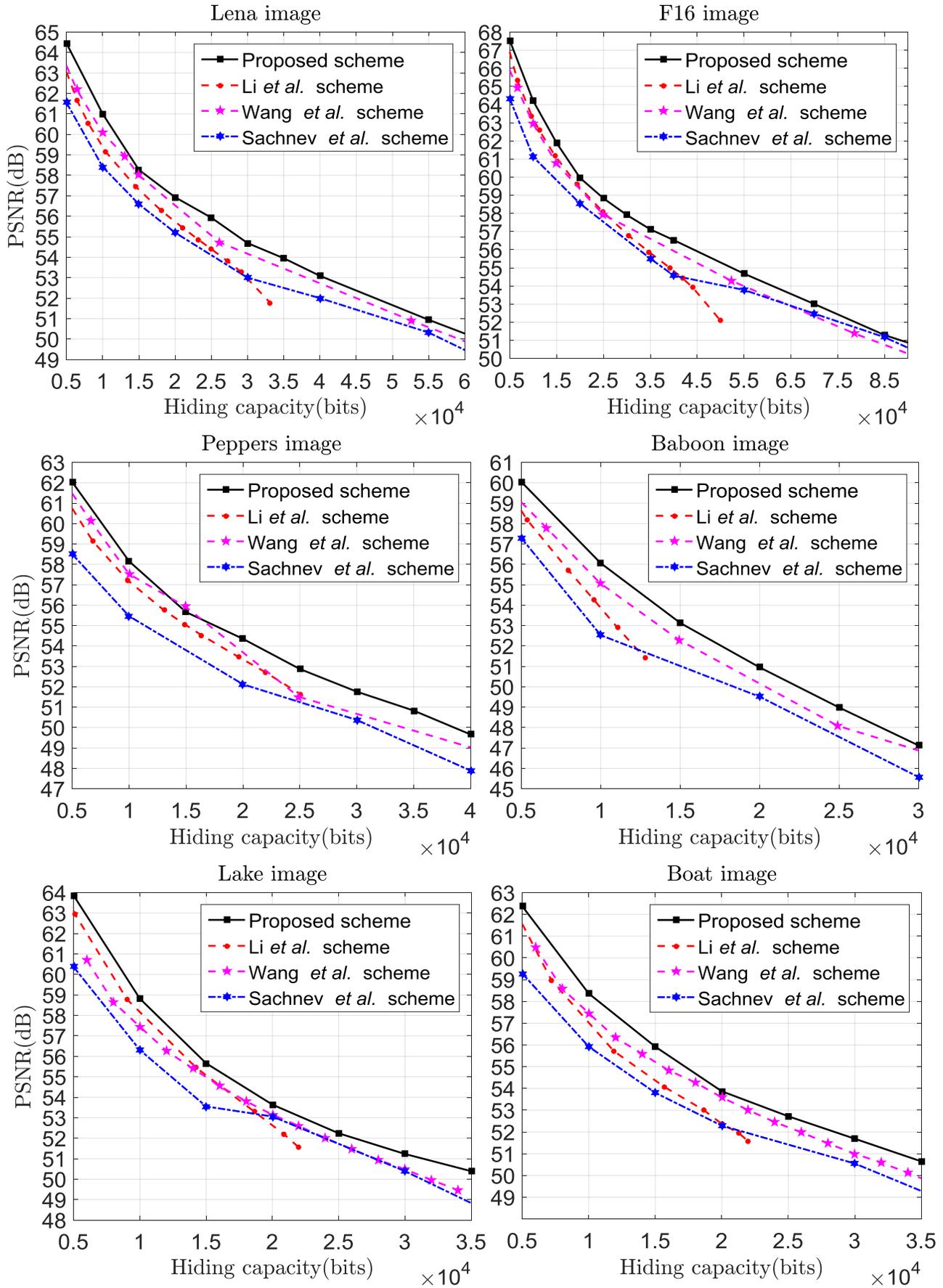

**Fig. 7** Comparing the performance of the proposed scheme with Li et al. [21], Wang et al. [23], and Sachnev et al. [5] schemes for 6 test images.

In order to demonstrate the importance of using $f_c = \overline{LD}$ in (11) instead of using $f_c = LD_{center}$ in (9), we provide the results of PSNR for different ECs for both cases. The calculated PSNR is listed in Table 3, where the role of side cells to improve marked image quality for specified ECs is not unassailable. This improvement is more significant for low ECs. Also, this improvement on smoother images, such as F16, is more considerable than rougher ones, like Baboon. For example in F16 at EC = $0.5 \times 10^3$ bits, we achieve more than 2 dB improvement in marked image quality using side cells along with center one, i.e. using $f_c = \overline{LD}$. For Lena image as a middle one in the term of roughness or smoothness, the improvement is, respectively, more than 1.5 dB and 0.7 dB for EC = $1 \times 10^3$ and EC = $5 \times 10^3$ bits.

Table 4 demonstrates the effect of applying the HIA and SPEP tools to the test images. The comparison of results for $f_c = \overline{LD}$ in Table 3 and HIA and SPEP in Table 4 reveals that they are more effective to improve marked image quality at lower ECs. Furthermore, the effect of SPEP to improve marked image quality is more significant than HIA for the smoother images. The application of HIA is more effective to improve PSNR for rougher images. For example, at ECs less than $20 \times 10^3$ (Table 4), there is no positive effect in using HIA for F16 and Lake. Nevertheless, for Baboon, the HIA improves quality of the marked image at least 1dB at these ECs. This improvement for the Peppers is more than 2 dB. For the Boat and Lena, the impact of HIA over SPEP or vice versa is not, significantly, different at many ECs. However, using HIA in company with SPEP, one can observe from Table 4 that the PSNR improvement is noticeable. For Boat at EC = $10^4$ bits, the use of SPEP and HIA together
achieves 1.12 dB and 0.62 dB enhancement, respectively, over the use of SPEP and HIA alone. This improvement for the majority of cases is more than employing HIA and SPEP alone.

In Fig. 7, a comparison between the proposed scheme and the schemes in [5], [21], [23] is performed for the test images. In our scheme, we use HIA with the two novel sorting methods. It can be seen that the image quality of the marked image is significantly improved, especially at lower ECs. Comparing with [5], the improvement of PSNR for the proposed scheme at some ECs is even more than 3 dB.

Reference [21] proposes a RDH scheme based on two-dimensional difference histogram modification and difference-pair-mapping (DPM). This scheme provides a good improvement in the marked image quality, especially at low ECs. In [21], the slope of the PSNR reduction is more than ours and the other aforementioned schemes. The proposed algorithm presents superior results to the scheme in [21], especially at higher ECs. The slope of the PSNR reduction for the scheme of [23] is approximately the same as ours. The improvement over this scheme for the test images is quite tangible.

As shown in Fig. 7, it seems our scheme poses no significant improvement in terms of PSNR over the other schemes at some ECs. Depicting Fig. 8 for F16 image, we show that the proposed scheme improves considerably the hiding capacity over the other schemes for a fixed PSNR. It can be seen from Fig. 8 that the minimum and maximum relative increase of EC by our scheme are nearly 6.5% and 22% over the scheme in [21] for PSNR = 61 dB and 64.2 dB respectively. Consequently, it is evident that our proposed scheme own enough merits over the previous ones.

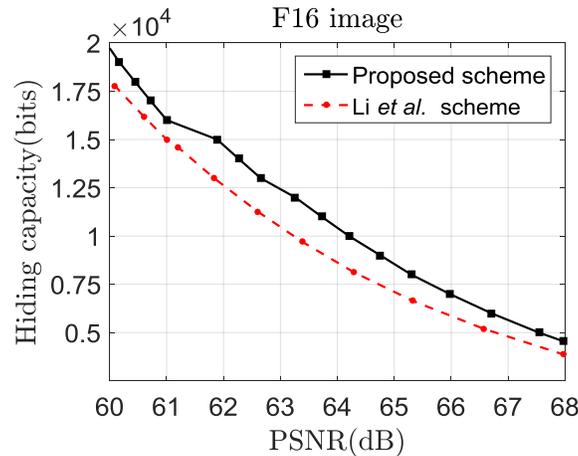

**Fig. 8** Comparing the hiding capacity of the proposed scheme with Li et al. [21] for F16.

## 5 Conclusion

We propose two novel sorting methods in order to classify cells into different cell frequency bands and divide a cell frequency band into different sub-bands. In this way, smooth pixels are, better, identified and more predictability for data hiding is achieved. In addition, we propose a hiding intensity analysis (HIA) to produce optimal PE that in turn, makes less distortion to the marked image. The comparison of the proposed scheme with other state of the art schemes confirms that our scheme improves significantly the quality of the marked image in terms of PSNR, especially for lower ECs. Moreover, our scheme poses considerable improvement in hiding capacity for a specific level of PSNR comparing to the other schemes.